\documentclass[superscriptaddress,prl,twocolumn,nofootinbib]{revtex4}
\usepackage{pdfpages}
\usepackage{epstopdf}
\usepackage{color}
\usepackage{graphicx}
\usepackage[justification=centerlast]{caption}
\usepackage{subcaption}
\usepackage{amsmath}
\usepackage[normalem]{ulem}

\newcommand{\iu}{\mathrm{i}}

\newcommand{\ket}[1]{|#1\rangle}

\usepackage{hyperref} 
\hypersetup{bookmarks=true,colorlinks=true,linkcolor=blue,citecolor=blue}

\begin{document}

\title{Integrated Silicon Photonics for High-Speed Quantum Key Distribution}

\author{P. Sibson}
\email[]{philip.sibson@bristol.ac.uk}
\author{J. E. Kennard}
\affiliation{Quantum Engineering Technology Labs, H. H. Wills Physics Laboratory \& Department of Electrical and Electronic Engineering, University of Bristol, Merchant Venturers Building,  Woodland Road, Bristol, BS8 1UB, UK}

\author{S. Stanisic}
\affiliation{Quantum Engineering Technology Labs, H. H. Wills Physics Laboratory \& Department of Electrical and Electronic Engineering, University of Bristol, Merchant Venturers Building,  Woodland Road, Bristol, BS8 1UB, UK}
\affiliation{Quantum Engineering Centre for Doctoral Training,  H. H. Wills Physics Laboratory \& Department of Electrical and Electronic Engineering, University of Bristol, Merchant Venturers Building,  Woodland Road, BS8 1UB, UK}

\author{C. Erven}
\author{J. L. O'Brien}
\author{M. G. Thompson}
\affiliation{Quantum Engineering Technology Labs, H. H. Wills Physics Laboratory \& Department of Electrical and Electronic Engineering, University of Bristol, Merchant Venturers Building,  Woodland Road, Bristol, BS8 1UB, UK}

\begin{abstract}
Integrated photonics offers great potential for quantum communication devices in terms of complexity, robustness and scalability. Silicon photonics in particular is a leading platform for quantum photonic technologies, with further benefits of miniaturisation, cost-effective device manufacture and compatibility with CMOS microelectronics. However, effective techniques for high-speed modulation of quantum states in standard silicon photonic platforms have been limited. Here we overcome this limitation and demonstrate high-speed low-error quantum key distribution modulation with silicon photonic devices combining slow thermo-optic DC biases and fast (10~GHz bandwidth) carrier-depletion modulation. The ability to scale up these integrated circuits and incorporate microelectronics opens the way to new and advanced integrated quantum communication technologies and larger adoption of quantum-secured communications.
\end{abstract}

\maketitle


\begin{figure*}
	\centering
	\includegraphics[trim = 8mm 7mm 5mm 8mm, clip, width=\textwidth]{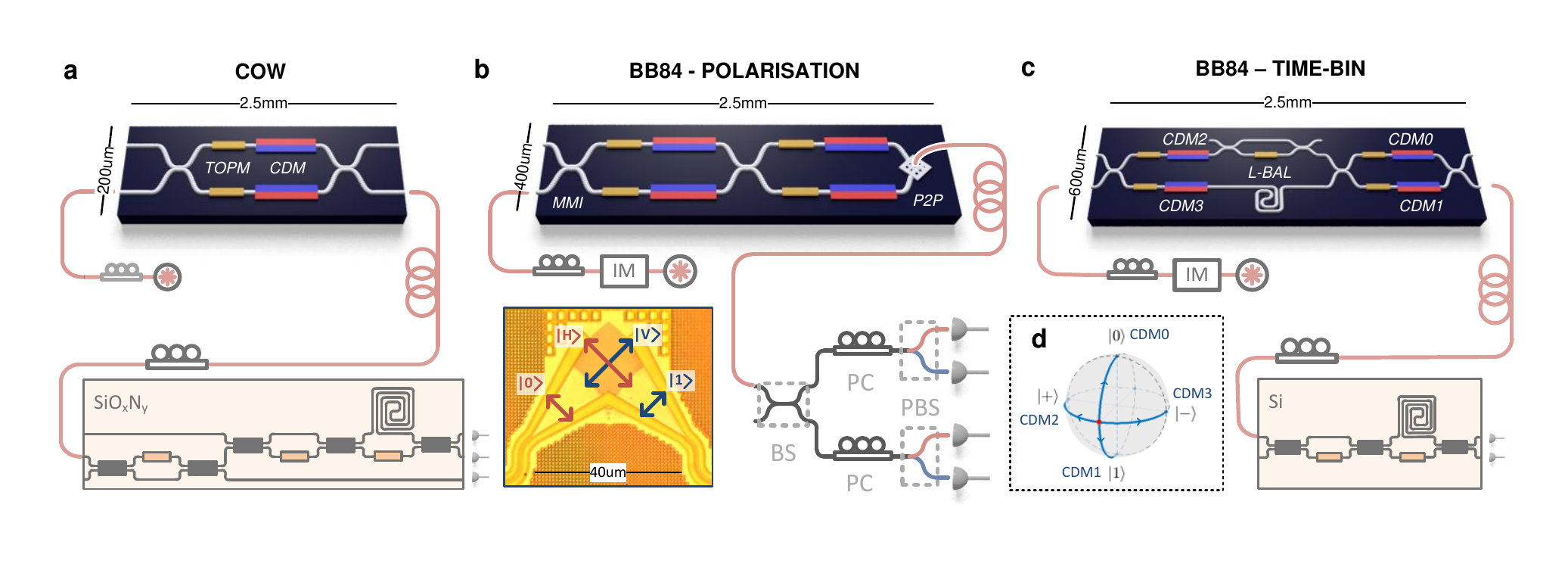}
	\caption{\textbf{Integrated Silicon Photonic Devices for QKD:} \textbf{(a)} Coherent-One-Way: A balanced Mach-Zehnder interferometer (MZI) comprising two multi-mode interference (MMI) devices acting as beam splitters, with phase modulation from thermo-optic phase modulators (TOPM) and carrier-depletion modulators (CDM) allows for the encoding of quantum information in path, or pulse modulation. \textbf{(b)} Polarisation encoded BB84: Combining the two paths of the MZI with a two-dimensional grating coupler allows for the conversion from path encoded information to polarisation encoded information (P2P), suitable for communication in free-space. \textbf{(c)} Time-bin encoded BB84: An unbalanced asymmetric MZI (AMZI) allows for encoding in time by temporally separating weak coherent pulses into two time intervals using an on-chip delay of 1.5~ns. The extra loss this incurs is balanced by an MZI used as a tunable beam splitter on the opposing arm. The last beam splitter in the AMZI is replaced with another MZI that allows for the selection of time-bin $\ket{0}$ and $\ket{1}$ states. DC offsets are provided by the TOPMs and fast modulation by four CDMs. \textbf{(d)} An illustrated Bloch sphere highlighting the DC offset at $\ket{+\iu}$ (red dot), set by the TOPMs. Each CDM is only required to modulate up to $\pi/2$ to permit the encoding of each BB84 state.}
	\label{fig:SCHE}
\end{figure*}

\noindent Quantum technologies are rapidly developing and have the potential to revolutionise the fields of computing and telecommunications. They have major implications for the security of many of our conventional cryptographic techniques, which are known to be insecure against a quantum computer \cite{QKDREVIEW}. Fortunately, quantum key distribution (QKD) provides a highly secure approach to sharing random encryption keys by transmitting single photons \cite{Scarani2009}. Although QKD has advanced from simple proof-of-principle experiments towards robust long-term demonstrations \cite{Yoshino:13,Korzh:13,Dixon:15,sasaki2015quantum}, it has still not obtained wide-scale adoption.

Integrated photonics provides a stable, compact, and robust platform to implement complex photonic circuits amenable to mass-manufacture, and therefore provides a compelling technology for optical quantum information devices \cite{INTREVIEW}. Silicon photonics, in particular, is a leading platform for quantum photonic technologies with the promise of high density integration, mature fabrication processing, and compatibility with microelectronics \cite{lim2014review}. Silicon has been used to demonstrate sources of quantum light, manipulation and transmission of quantum information, and integration with single photon detectors \cite{Silverston2016Silicon,Pernice2012,IntegratedSNSPDS}. It has also been used in classical computing and communications for modulation, transceivers \cite{doerr2015silicon}, and a recent demonstration of optical interconnects alongside electronic microprocessor technology \cite{sun2015single}.

The appeal of this platform has led to integrated photonic technologies increasingly being deployed in the development of practical QKD systems. Demonstrations include integrated ``client'' chips for Reference-Frame-Independent QKD \cite{RFI}, planar waveguide components in transmitters and receivers \cite{tanaka2012high}, and chip-to-chip QKD using GHz clocked indium phosphide transmitters and silicon oxynitride receivers \cite{sibson2015chip}. 

However, high-speed modulation of quantum states in standard silicon photonic fabrication has been limited. With no natural electro-optic non-linearity, many silicon quantum photonic experiments instead utilise slow thermo-optic phase modulators (TOPMs) for high-fidelity state preparation. Carrier injection or carrier depletion modulators (CDMs) offer high-speed operation, but incur phase dependent loss and saturation \cite{reed2010silicon}, which are detrimental in quantum applications where state preparation has stringent requirements. 

Here we show an approach to overcome the limitations of saturation and phase dependent loss of high-speed CDMs in standard silicon photonic fabrication.
First we describe a combination of slow, but ideal, TOPMs alongside fast, but non-ideal, CDMs utilised for QKD state preparation at GHz speeds. We then use this technique to demonstrate three implementations of high-speed low-error QKD (Figure~\ref{fig:SCHE}): chip-to-chip Coherent One Way (COW) QKD, polarisation encoded BB84, and time-bin encoded BB84 \cite{BB84} state preparation. We achieve estimated asymptotic secret key rates of up to 916~kbps and quantum bit error rates (QBER) as low as 1.01\% over 20~km of fibre, experimentally demonstrating the feasibility of high-speed QKD integrated circuits based on standard silicon photonic fabrication.

\begin{figure}[t]
	\centering
	\begin{subfigure}{0.5\textwidth}
			\includegraphics[trim = -5mm 5mm 0mm 5mm, clip, width=\textwidth]{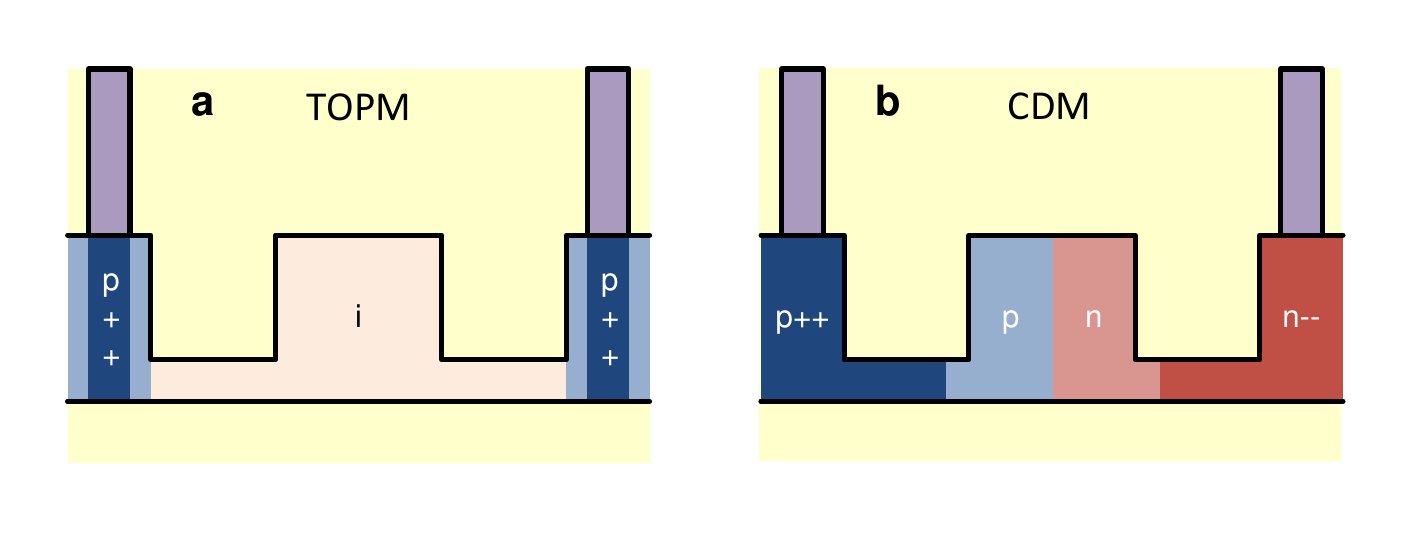}
	\end{subfigure}
	\begin{subfigure}{0.5\textwidth}
		\includegraphics[trim = 10mm 0mm 10mm 10mm, clip, width=\textwidth]{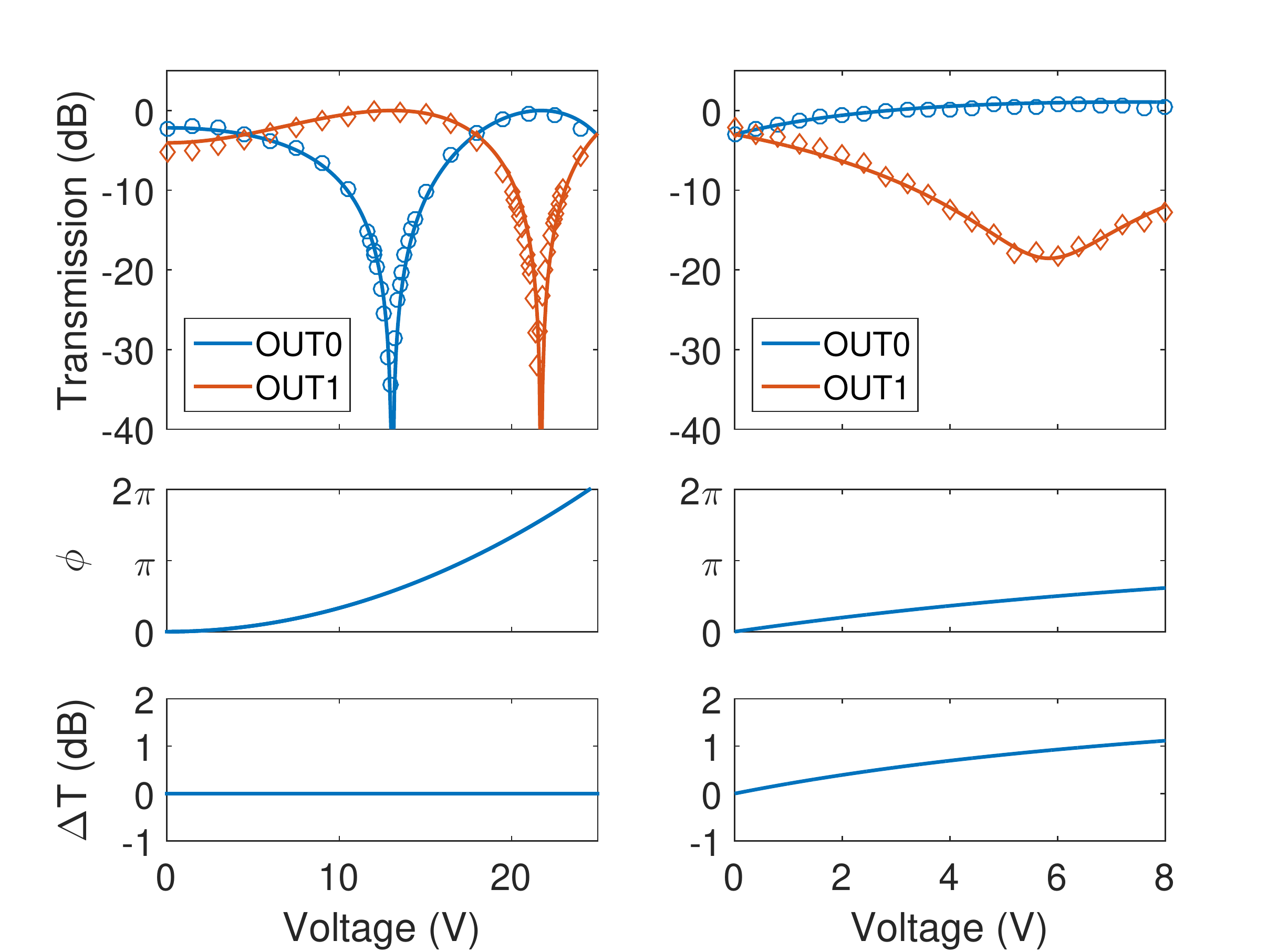}
	\end{subfigure}
	\caption{\textbf{Thermo-optic and Carrier-depletion Phase Modulation in Silicon Photonics} fabricated with standard doping processes \cite{absil2015imec}. \textbf{(a)} Cross section of the thermo-optic phase modulation waveguide with \emph{p}++ doping in the waveguide slab and intrinsic (\emph{i}) silicon waveguide core, followed by the power measured at the two outputs of an MZI, the fitted quadratic phase ($\phi$) relationship, and the change in transmission ($\Delta \mathrm{T}$) as a function of the applied voltage (V). \textbf{(b)} Cross section of the carrier depletion phase modulator with \emph{p} and \emph{n} doping in the waveguide core, followed by the power measured at the two outputs of an MZI (with an additional TOPM providing a $\pi/2$ offset or initially equal intensity outputs), the fitted phase ($\phi$) relationship illustrating saturation, and the change in transmission ($\Delta \mathrm{T}$) as a function of the applied voltage (V).}
	\label{fig:CARRIER_TOPS}
\end{figure}

\vspace{2.5mm}
Figure~\ref{fig:SCHE} shows the three different silicon photonic devices which use a combination of TOPMs and CDMs to prepare and modulate QKD states. As illustrated in Figure~\ref{fig:CARRIER_TOPS}~(a), the TOPMs allow for almost ideal state preparation with a phase relationship proportional to the square of the voltage ($V^{2}$) and no change in transmission ($\Delta \mathrm{T}$). The TOPMs were designed in silicon-on-insulator using doped resistive heating in the waveguide slab. This design provides ohmic electrical characteristics ($\sim$6.14~k$\Omega$ for a 150~$\mu$m length) and a $2\pi$ voltage of $\sim$24~V.

The TOPMs are limited in bandwidth ($\sim$kHz) and therefore inappropriate for the fast modulation required in communications. Silicon also has no natural $\chi^{(2)}$ non-linearity, and therefore the high-speed electro-optic effect cannot be used. An approach to overcome this in standard silicon photonic platforms is to use doped waveguide sections as shown in Figure~\ref{fig:CARRIER_TOPS}~(b), and employ carrier injection or depletion techniques.

Carrier depletion modulation induces a phase by reducing the carriers occupying the region which overlaps with the optical mode. This is achieved by reverse biasing a \emph{p-n} junction formed by doping \emph{p} and \emph{n} regions in the core of the silicon waveguide. The depletion of carriers in the waveguide decreases the absorption of the waveguide and induces an optical phase change \cite{fang2012recent}; however, this effect saturates as the waveguide becomes fully depleted of carriers. Figure~\ref{fig:CARRIER_TOPS}~(b) plots the extracted relationship between voltage, phase, and transmission for a 1.5~mm modulator ($\sim$5~dB loss at 0~V), illustrating that the induced phase saturates below $\pi$ over 8~V while the transmission continues to increase.

The phase-dependent loss characteristics and saturation of the CDMs can severely reduce operational fidelity. This is especially damaging for quantum applications where requirements are often more stringent than many classical applications. Here we describe a combination of thermal and carrier depletion modulators to minimise this effect. By using the TOPMs to bias the circuits in a favourable operating regime, and limiting the modulation depth required for each individual CDM, we mitigate these negative characteristics for both pulse modulation and state preparation.

\vspace{2.5mm}
Utilising this technique, we demonstrate the pulse modulation of coherent light using a Mach-Zehnder interferometer (MZI) including both TOPMs and CDMs, as illustrated in Figure~\ref{fig:SCHE}~(a). The TOPMs provide a DC offset to minimise one of the MZI output intensities. Consequently, small changes in one of the CDMs' phases will cause a large change in intensity of this output arm with a high extinction ratio of $\sim$25~dB without requiring a full $\pi$ phase change. The -3~dB bandwidth of the CDM changes with the biasing conditions, and was estimated at $\sim$10~GHz and produced $\sim$175~ps full-width-half-maximum (FWHM) pulses. This allows operation of the devices for use in the Coherent-One-Way (COW) QKD protocol.

COW is a distributed-phase-reference scheme \cite{stucki2005fast} that transmits pulses in pairs, encoding $\ket{0}$ in the first bin and $\ket{1}$ in the second. Here we modulate an external CW laser (1550~nm) to generate pulses in these time-bins and generate a key from the unambiguous time of arrival of the single photons in each time-bin pair. Security of the channel is maintained by measuring the visibility from interfering successive photon pulses, when the pattern of state $\ket{1}$ followed by state $\ket{0}$ occurs, at the receiver \cite{branciard2008upper}. A decoy state, with photon pulses in each time-bin ($\ket{0}$ and $\ket{1}$) is included to increase the probability of occupied successive pulses allowing a more accurate measurement of the interference, and to detect photon-number-splitting attacks.

We perform chip-to-chip COW QKD using the integrated SiO$_x$N$_y$ receiver device from Ref.~\cite{sibson2015chip} as illustrated in Figure~\ref{fig:SCHE}~(a). The receiver uses a reconfigurable thermo-optic MZI to route a larger proportion of the input signal directly to a single photon detector for key generation. The fibre-coupled superconducting nanowire single photon detectors are biased to $\sim$40\% system detection efficiency and dark count rates of $\sim$500~cps per detector. A smaller proportion of the signal is routed to an asymmetric MZI (AMZI) with a 580~ps delay line for the visibility measurement to verify to security of the channel.

\vspace{2.5mm}
Extending this approach also allows the state preparation of the four BB84 states \cite{BB84}. Figure~\ref{fig:SCHE}~(b) illustrates slow TOPMs inside and outside an MZI that apply a static phase offset, while we use the non-ideal but fast CDMs in both arms to apply small state-dependent modulations. As shown in Figure~\ref{fig:SCHE}~(d), we start by preparing the state $\ket{+\iu}$ (red dot on the Bloch sphere), using the TOPMs as a DC offset, with near unit fidelity. We then use four non-ideal CDMs (two inside and two outside of the MZI) to prepare any of the four BB84 states $\{\ket{0}, \ket{1}, \ket{+}, \ket{-}\}$ by shifting each of the fast CDMs by $\pi/2$ respectively. This approach limits each CDM to a $\pi/2$ phase-shift thus minimising the phase dependent transmission across all four BB84 states and allowing state preparation with non-ideal phase modulators.

The technique can be applied to polarisation encoding as illustrated in Figure~\ref{fig:SCHE}~(b). We couple an intensity modulated (IM) laser to launch 175~ps FWHM optical pulses into a path encoded MZI. Applying static biases to the TOPMs in the MZI prepares the $\ket{+\iu}$ state, and using each of the CDMs generates one of the four BB84 states as above. Finally this path encoded state is combined on a 2D grating coupler that acts as a path-to-polarisation converter \cite{wang2016chip}.

A fibre-based receiver [shown under the transmitter chip in Figure~\ref{fig:SCHE}~(b)] was constructed to decode and measure the output polarisation states. It consisted of a 50:50 beam splitter (BS) to passively choose the measurement basis, with one arm connected to a polarisation beam-splitter (PBS) to measure $\ket{0}$ and $\ket{1}$, and one arm coupled to a PBS to measure $\ket{+}$ and $\ket{-}$. Polarisation controllers (PC) were then used to set the measurement bases in each arm. Finally, the outputs were coupled to superconducting nanowire single photon detectors.

\begin{figure*}
	\centering
	\includegraphics[trim = 53mm 0mm 44mm 0mm, clip, width=\textwidth]{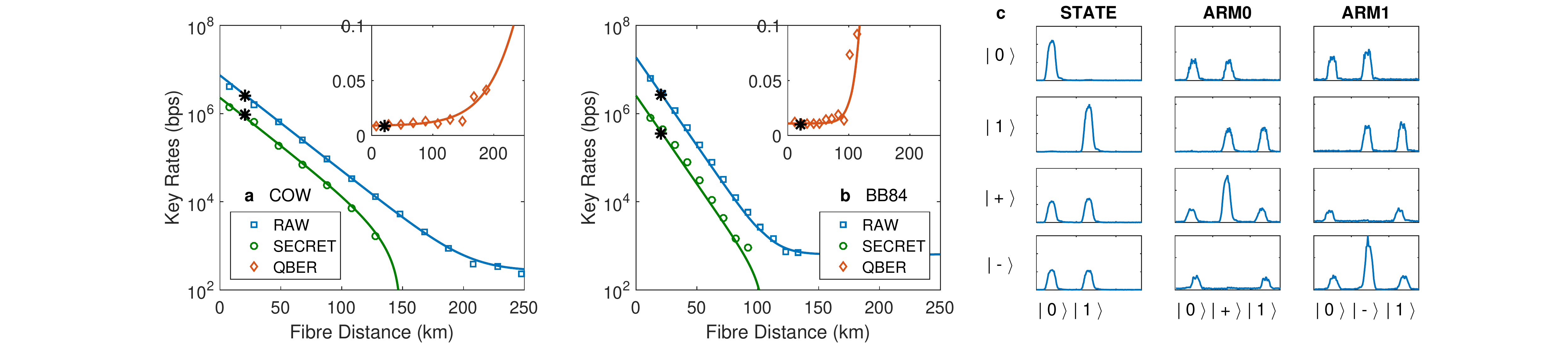}
	\caption{\textbf{Estimated Secret Key Rates:} The main data sets (squares, circles, and diamonds) were collected by emulating a quantum channel with the use of a variable optical attenuator and assuming standard fibre losses; however, the data shown with asterisks was collected using a 20~km fibre spool as the quantum channel. \textbf{(a)} Raw and secure key rates using the chip to implement the COW QKD protocol, as illustrated in Figure~\ref{fig:SCHE}~(a). The system operates with a 1.72~GHz clock-rate with a QBER of 1.01\% and estimated secure key rate of 916~kbps over a 20~km fibre. \textbf{(b)} Raw and secure key rates using the chip to produce polarisation encoded BB84 states, as illustrated in Figure~\ref{fig:SCHE}~(b).  We measure a low QBER of 1.1\%  while the transmitter is operated with a 1~GHz clock-rate which yields an estimated secure key rate of 329~kbps over a 20~km. \textbf{(c)} Histogram measurements of the time-bin encoded BB84 state preparation and measurement, as illustrated in Figure~\ref{fig:SCHE}~(c). }
	\label{fig:RATES}
\end{figure*}

\vspace{2.5mm}
Polarisation encoded systems are commonly used in free-space links owing to the non-birefringent nature of the atmosphere. However, in fibre-based networks the birefringence of the fibre induces a time-dependent polarisation rotation on any transmitted qubits, requiring active compensation to maintain reference frame alignment \cite{xavier2009experimental}. Instead, encoding qubits in time and phase, so called time-bin encoding, when transmitting over fibre optics is a much more stable degree of freedom. Here we show the combined TOPM and CDM approach for state preparation described above can be translated to the equivalent time-bin encoded device.

A schematic of the integrated circuit used to produce time-bin qubits is illustrated in Figure~\ref{fig:SCHE}~(c). An external laser (1550~nm) is intensity modulated (IM) to generate weak coherent pulses (350~ps FWHM), which are coupled into the silicon chip containing an AMZI with a 1.5~ns delay line to separate photon pulses into one of two time-bins. Here, $\ket{0}$ and $\ket{1}$ is encoded by a photon in the first or second time-bin, respectively, and $\ket{+}$ and $\ket{-}$ encoded by a photon in a superposition of the first and second time-bin with $0$ and $\pi$ relative phase, respectively. In future designs, the external LiNbO$_3$ IM can be replaced with an on-chip MZI as demonstrated in the COW protocol, as they showed comparable performance with extinction ratios of 25-30~dB. Although the modulation bandwidth and loss performance of the on-chip MZI is inferior to the external LiNbO$_3$ IM (10~GHz rather than 40~GHz bandwidth, and 5~dB loss rather than 4~dB insertion loss), they are still sufficient for QKD and gain the important advantage of monolithic integration in a standard silicon photonic platform.

An MZI is included in the top arm and contains a TOPM to apply a loss balancing correction to ensure each time-bin contains equal photon amplitudes exiting the device. The inside of the AMZI includes TOPMs which can control the DC relative phase between the separate pulses and CDMs to alternate between the $\ket{+}$ and $\ket{-}$ states. The final beam splitter of the AMZI is replaced with a further MZI, with the TOPMs biasing an equal superposition of the two time-bins, and the two CDMs used to select the first or second time-bin coupled into a single fibre output.

A second identical chip [pictured at the bottom right of Figure~\ref{fig:SCHE}~(c)] is used as a receiver circuit with a matched AMZI used to passively select basis measurements before the output is fibre coupled and measured with superconducting nanowire single photon detectors. The phase decoding AMZI overlaps successive time-bins creating three possible time-slots within which to detect photons. Phase information is interfered in the middle time-slot allowing measurements in the $\{\ket{+}, \ket{-}\}$ basis, whereas time of arrival information in the first and third time-slots allows measurement in the $\{\ket{0}, \ket{1}\}$ basis.

\vspace{2.5mm}
Figure~\ref{fig:RATES} illustrates the results from our three implementations of high-speed low-error QKD with integrated silicon photonics. Figure~\ref{fig:RATES}~(a) shows the raw and secret key rates, and the QBER from the system performing the COW QKD protocol. Here, pulse modulation provides 175~ps FWHM pulses with a high $\sim$25~dB extinction ratio between bright and empty pulses. The system operates with a 1.72~GHz clock-rate (or 0.86~GHz system-rate, as one state is sent every two clock cycles) with a QBER of 1.01\% and estimated asymptotic secure key rate of 916~kbps over a 20~km fibre, following the upper-bound security proof against collective attacks of Branciard \emph{et al.} \cite{branciard2008upper}.

For the polarisation and time-bin BB84 QKD protocols, we model and fit the data from our TOPMs and CDMs in Figure~\ref{fig:CARRIER_TOPS}~(a) and~(b), resulting in an expected state preparation fidelity of 99.5\%, which yields an expected QBER of $\le$1.1\%. This is equivalent to a 19.5~dB extinction ratio between measuring the state $\ket{\psi}_i$ and its orthogonal counterpart $\ket{\bar{\psi}}_i$ (e.g. measuring $\ket{1}$ when preparing $\ket{0}$ or measuring $\ket{-}$ when preparing $\ket{+}$).

Figure~\ref{fig:RATES}~(b) shows the measured raw and secret key rates as well as the QBER from operating the transmitter for polarisation encoding and using the passive fibre-based receiver detection scheme described. We measure a low QBER of 1.1\%  while the transmitter is operated with a 1~GHz clock-rate which yields an estimated asymptotic secure key rate of 329~kbps over a 20~km fibre using a non-phase randomised weak coherent BB84 security proof without decoy states against general attacks \cite{lo2007security}. The rate and maximum secure distance could be increased drastically with the addition of an extra integrated intensity modulator to produce decoy states \cite{2005Decoyproof}.

Finally, measurements of the time-bin encoded states, using a second identical chip as the receiver circuit as described above, are shown in Figure~\ref{fig:RATES}~(c). Analysing these measurements, we observe a low QBER of $\sim$2.1\% for this proof-of-principle demonstration of time-bin encoded BB84 state preparation. 
Future systems will benefit from a dedicated low-loss silicon receiver circuit by minimising the fibre-to-chip coupling loss (currently $\sim$-4.5~dB) and reducing the loss incurred in the on-chip delay; e.g., reducing the 1.5~ns delay used here to the 600~ps used in COW would increase transmission by $>$3~dB. As we implement a passive basis selection scheme there is not a need for fast modulation on the receiver, and removing the \emph{p-n} carrier depletion modulators would further decrease absorption by $\sim$5~dB per modulator.

\vspace{2.5mm}
In conclusion, this work experimentally demonstrates the feasibility of high-speed QKD transmitters in CMOS-based silicon photonic integrated circuits. In particular, we show an approach to overcome the problems of high-fidelity state preparation when using non-ideal fast modulation in standard silicon photonic fabrication. Using a combination of slow, but ideal TOPMs alongside high-bandwidth ($\sim$10~GHz), but non-ideal, CDMs we demonstrate QKD state preparation and pulse modulation. We show three successful implementations: time-bin encoded BB84 state preparation and measurement, polarisation encoded BB84 (1~GHz clock-rate, 1.1\% QBER, 329~kbps estimated asymptotic secret key rate), and pulse modulation for COW QKD (1.72~GHz clock-rate, 1.01\% QBER, 916~kbps estimated asymptotic secure key rate) over a 20~km fibre link.

The modulation bandwidth and estimated secret key rates generated in this demonstration are comparable to previous integrated photonic QKD demonstrations \cite{sibson2015chip}, but with less functionality combined on a single device and currently relying on off-chip laser sources. Future generations of silicon based chips will benefit from the recent developments of low-loss couplers \cite{cardenas2014high}, low-loss delay lines \cite{cardenas2009low}, integrated laser sources \cite{liang2010recent}, and integrated single photon detectors \cite{marsili2013detecting}, allowing high-performing monolithically integrated transmitter and receiver devices. Performance could be further improved by combining the functionality demonstrated, such as integrated pulse modulation, intensity modulators for decoy-state \cite{2005Decoyproof}, and attenuation calibration, and by increasing the rates beyond the 10~GHz bandwidth demonstrated here \cite{reed2010silicon}. The ease of fabrication and availability of silicon photonics (in comparison with other integrated photonic platforms such as InP and SiO$_x$N$_y$) will also open routes to mass-manufacture using standard CMOS fabrication tools and foundries.

Future demonstrations will require focus on the complete system for autonomous QKD operation deployed in telecommunication networks. This includes the development of real-time basis and bit selection using quantum random numbers, active basis alignment, appropriate error reconciliation, privacy amplification, and the use of finite-key analysis to qualify the security.

Ultimately, integrated silicon photonics will allow the manufacture of quantum communication chips with electronic and photonic processing on a single monolithic device and the reduced footprint will enable further multiplexing, complexity and operation with single photon detection. The ability to scale up these integrated circuits and incorporate microelectronics opens the way to new and advanced integrated quantum communication technologies and larger adoption of quantum-secured communications.

\vspace{2.5mm}
During the preparation of this manuscript, the authors became aware of a complementary demonstration by Ma \emph{et al.} \cite{Ma2016} in silicon using carrier injection \emph{p-i-n} diode phase-shifters to encode BB84 states in polarisation and micro-ring resonators for pulse modulation.


\vspace{-0.5mm}
\section*{Acknowledgments}
The authors thank IMEC for the fabrication of the integrated photonic silicon devices and LioniX for the fabrication of the silicon oxynitride receiver. 

This work was supported by the Engineering and Physical Sciences Research Council (EPSRC), The European Research Council (ERC), FP7 Action: Beyond the Barriers of Optical Integration (BBOI), and the UK Quantum Communications Hub. J.L.O'B. acknowledges a Royal Society Wolfson Merit Award and a Royal Academy of Engineering Chair in Emerging Technologies. M.G.T acknowledges fellowship support from the Engineering and Physical Sciences Research Council (EPSRC, UK).


\bibliographystyle{ieeetr}
\bibliography{SI_QKD_REFS}

\end{document}